\documentclass[oneside,british]{amsart}
\usepackage[T1]{fontenc}
\usepackage[latin9]{inputenc}
\usepackage{float}
\usepackage{amsthm}
\usepackage{graphicx}
\usepackage{esint}
\usepackage[authoryear]{natbib}
\usepackage{url}
\usepackage{algorithmic}

\floatstyle{ruled}
\newfloat{algorithm}{tbp}{loa}
\floatname{algorithm}{Algorithm}

\numberwithin{equation}{section} %% Comment out for sequentially-numbered
\numberwithin{figure}{section} %% Comment out for sequentially-numbered
\theoremstyle{plain}

\makeatother

\usepackage{babel}

\begin{document}

\title{Fast simulation of truncated Gaussian distributions}

\author{Nicolas Chopin, ENSAE-CREST}
\begin{abstract}
  We consider the problem of simulating a Gaussian vector $X$,
  conditional on the fact that each component of $X$ belongs to a finite
  interval $[a_{i},b_{i}]$, or a semi-finite interval
  $[a_{i},+\infty)$. In the one-dimensional case, we design a
  table-based algorithm that is computationally faster than
  alternative algorithms. In the two-dimensional case, we design an
  accept-reject algorithm. According to our calculations and our
  numerical studies, the acceptance rate of this algorithm is bounded
  from below by $0.5$ for semi-finite truncation intervals, and by
  $0.47$ for finite intervals. Extension to 3 or more dimensions is
  discussed.

\end{abstract}
\maketitle

\section{Introduction}

Let $X=(X_{1},\ldots,X_{d})$ be a $d-$dimensional Gaussian vector with
mean $\mu$ and covariance matrix $\Sigma$, and let $[a_{i},b_{i}]$
be $d$ intervals, where $b_{i}$ may be either a real number or $+\infty$.
The distribution of $X$, conditional on the event that $X_{i}\in[a_{i},b_{i}]$,
$i=1,\ldots,d$, is usually called a truncated Gaussian distribution
\citep[Chap. 13]{UniContDist1}. Without loss of generality, one may
assume that $\mu=0$, and that $\Sigma$ has unit diagonal elements. 

Numerous statistical algorithms rely on intensive simulation of
truncated Gaussian distributions. In particular, several Bayesian
models generate full conditional distributions of this type, either
directly or through a data augmentation representation
\citep{TannerWong}. Thus, the corresponding Gibbs samplers (or more
generally Markov chain Monte Carlo algorithms) draw repetitively from
truncated Gaussian distributions.  Examples include linear regression
models with ordered parameters \citep{ChenDeely} or applied to
truncated data \citep{GelSmiLee}, probit models \citep{Chib},
multinomial probit models
\citep{Chib,McculloghRossi:94,nobile:probit}, multivariate probit
models \citep{ChibGreen}, multiranked probit models \citep{LinaDella},
tobit models \citep{Chib:tobit}, models used in spectroscopy
\citep{GulamFitzAnd}, copula regression models \citep{PittChanKohn},
among others.

To understand how intensive such MCMC algorithms can be, consider the
problem of sampling the posterior distribution of a multinomial probit
model with $n$ observations and $p$ alternatives. A solution is to 
 perform $T$ iterations of the Gibbs sampler of
\citet{McculloghRossi:94}, but this requires the generation of $Tnp$
univariate truncated Gaussian variates, a number that may exceed
$10^{12}$ or even $10^{15}$ in difficult scenarios. Hence any
improvement with respect to the computational cost of simulating
univariate truncated Gaussian distributions may lead to important
savings. Another important aspect of such algorithms is that they
simulate \emph{only one} random variable from a given truncated
Gaussian distribution, that is, the parameters and the truncation
intervals $[a_{i},b_{i}]$ change every time a truncated Gaussian
variate is generated. Thus, we are interested in developing
specialised algorithms which are guaranteed to generate quickly one
random variable from the desired distribution, for all possible inputs
(i.e., parameters and truncation intervals). As a corollary, these
algorithms cannot afford a long set-up (initialisation) time, where
some exploration of the target density is performed in order to
improve performance; this type of initialisation is meaningful only
when one needs to simulate many variables from one fixed distribution,
and is not discussed in this paper. The algorithm we propose does
require a table set-up, but which is independent of the input
parameters.

The first part of this paper presents a table-based simulation algorithm
for univariate Gaussian distributions truncated to either a finite
interval $[a,b]$ or a semi-finite interval $[a,+\infty).$ In the
latter case, and given the truncation point $a$, our algorithm is
up to three times faster than alternative algorithms in our simulations;
see below for references. Our algorithm is inspired from the Ziggurat
algorithm of \citet{Marsaglia:ziggurat,Marsaglia:zig2000}, which
is usually considered as the fastest Gaussian sampler, and is also
very close to  \citet{ahrens1995one}'s algorithm. 

Another possible strategy for accelerating a Gibbs sampler is to
`block', i.e., to update jointly, two or more components of the
posterior density; this often strongly improves the mixing properties
of the algorithm.  In some of the aforementioned models, blocking
requires simulating multivariate truncated Gaussian variates. We
develop an accept-reject algorithm for simulating from bivariate
truncated Gaussian distributions. In all but one particular case for
finite intervals, we manage to prove formally that the acceptance rate is
bounded from below by $0.22$. Our numerical studies seem to indicate
that this bound is not optimal, and that the acceptance rate is
bounded from below by $1/2$ when the truncation intervals are
semi-finite, and by $0.477$ when they are finite. (This remains true
even when the correlation coefficient get close to $1$ or $-1$, that
is, in situations where MCMC blocking is particularly efficient.) In
the former case, we explain how to generalise this algorithm in some
situations to truncated Gaussian distributions of dimension $d$, with
the outcome that the acceptance rate is bounded from below by
$1/2^{d-1}$. Interestingly, some of the constants that must be 
pre-computed for our univariate algorithm can be re-used so as to
bypass part of the computations performed by our multi-dimensional
algorithms.

We note that independent variables from truncated Gaussian distributions
may also be obtained using the perfect samplers of \citet{PhilRob}
and \citet{FerFerGryn}, but, for small dimensions, these algorithms
are much more expensive than our approach, since each sample requires
running a Markov chain until some criterion is fulfilled. (According
to \citet{Hoermann2006}, \citet{PhilRob}  may not sample from the
correct distribution.)

The paper is organised as follows. Section 2 presents our algorithm
for simulating univariate truncated Gaussian variables. Section 3
presents a rejection algorithm for simulating bivariate Gaussian vectors,
the components of which are truncated to semi-finite intervals $[a_{i},+\infty)$.
Section 4 does the same thing for finite truncation intervals. Section
5 explains how to generalise the algorithms of Section 3 to three
or more dimensions. Section 6 concludes.

\section{One-dimensional case}

First, we consider  the problem of simulating a random variable $X$ from
a univariate Gaussian density truncated to $[a,+\infty)$: 

\begin{equation}
p(x)=\frac{\varphi(x)}{\Phi(-a)}I(x\geq a)\label{eq:tn}\end{equation}
for some truncation point $a$, where $\varphi$ and  $\Phi$ denote
respectively the unit Gaussian probability density and cumulative distribution
functions; $\varphi(x)=\exp(-x^{2}/2)/\sqrt{2\pi}$. The extension
to a finite truncation interval $[a,b]$ is explained in \S\ref{sub:finite}.

\subsection{Review of current algorithms}\label{sec:revi-curr-algor}

A convenient way to generate $X$ is to use the inverse transform
method: \begin{equation}
X=-\Phi^{-1}\left(\Phi(-a)U\right),\label{eq:goodinverse}
\end{equation}
where $U\sim U[0,1]$ is a uniform variate. Note that this expression
is equivalent to 
\begin{equation}
\label{eq:badinverse}
X=\Phi^{-1}\left(\Phi(a)+\left\{ 1-\Phi(a)\right\} U\right),
\end{equation}
but the latter expression is less stable numerically for large values
of $a$, because it is easier to approximate $\Phi^{-1}$ in the left
tail than in the right tail. In our experiments, \eqref{eq:badinverse}
generates~ ``inf'' values when $a>9.5$, while \eqref{eq:goodinverse}
generates~``inf'' values only when $a>37.5$.

As noted by \citet[Chap. 2]{glasserman2003monte}, the inverse
transform method seldom produces the fastest algorithms, but it has
appealing properties that may justify the increased cost in some
settings, in particular when used in conjunction with variance
reduction or quasi Monte Carlo techniques; see the same reference and
also e.g. \cite{blair1976rational} for an overview of fast methods for
evaluating $\Phi$ and $\Phi^{-1}$. We now focus on specialised
algorithms.

First, we recall briefly the rejection principle
(\citealp[e.g.][Chap. 2]{Devroye:book} or \citealp{hormann:book},
Chap. 2). Assume we know of a proposal density $q$ such that
 \[
p(x)\leq Mq(x)\] 
for some $M\geq1$, and all $x$ in the support of
$q$. Then a sample from $p$ can be obtained as follows: simulate
$X\sim q$, and accept the realisation $x$ with probability $p(x)/Mq(x)$; otherwise
repeat. The expected acceptance probability, a.k.a. the acceptance
rate, equals $1/M.$ It is important to choose $q$ so that a) $M$ is
small and b) simulating from $q$ is cheap.$ $

For $a\geq0$, \citet[p. 382]{Devroye:book} proposes a rejection
algorithm based on the proposal exponential density
$q(x)=\lambda\exp\left\{ -\lambda(x-a)\right\} $, for $x>a$, with
$\lambda=a$. The acceptance rate of this algorithm is
$a\exp(a^{2}/2)\Phi(a)$, which goes to zero as $a\rightarrow0$, so it
can be used only for $a\geq a_{0}$, with say $a_{0}=1$. For $a<a_{0}$,
one may use instead the following trivial rejection algorithm: repeat
$X\sim N(0,1)$ until $X\geq a$. \citet[p. 382]{Devroye:book} mentions
\citet{marsaglia:tn}'s algorithm, which has the same acceptance rate,
but is a bit more expensive. \citet{Geweke:tn} and
\citet{Robert:TruncGaussian} independently derive a rejection
algorithm for $a\geq0$, based again on $q(x)=\lambda\exp\left\{
  -\lambda(x-a)\right\} $, but with $\lambda=(a+\sqrt{a^{2}+4})/2$,
which is shown to give the optimal acceptance rate. For $a<0$, these
authors use the same trivial sampler as above. In principle, these
algorithms may be refined using ARS \citep{GilksWild}, see also
\citet{hormann1995rts} and \citet{evans1998rvg}, i.e., rejected points
are used to improve the proposal density (which is then piecewise
exponential). As explained in the introduction however, we are
interested in situations when only one random variate must be
generated (for a given value of $a$); hence, since the acceptance rate
of Devroye's and Geweke and Robert's algorithms are high enough, we do
not discuss this further.

The algorithm we propose in this paper is faster than these specialised
algorithms for two reasons: (a) its acceptance rate is higher, and,
in fact, is almost one, for most values of $a$; and (b) with high
probability, the only floating point operations that the algorithm
performs are 2 additions and 3 multiplications, whereas the aforementioned
algorithms computes a few logarithms and square roots.

\subsection{Principle of proposed algorithm\label{sub:principle}}

\begin{figure}
\begin{centering}
\includegraphics[scale=0.6]{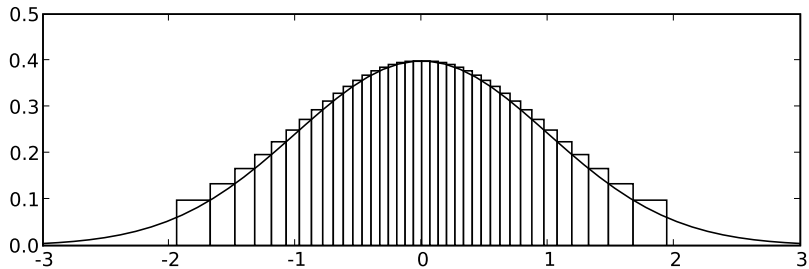}
\par\end{centering}

\caption{\label{fig:principle_zig}Plot of $N(0,1)$ density and the $2N$
vertical rectangles, for $N=20$.}

\end{figure}

For the sake of clarity, we consider first the simulation of a
non-truncated $N(0,1)$ density, and consider the extension to a
truncated density in next section. We do not claim, however, that this
algorithm is either interesting or novel in the non-truncated case,
see below for references. The principle of the algorithm is summarised
by Figure \ref{fig:principle_zig}. The proposal distribution consists
of $2N+2$ regions: $2N$ vertical rectangles of \emph{equal area}, and
two Gaussian tails of the same area. For rectangle $i$, $i=-N,\ldots,N-1$, let
$[x_{i},x_{i+1}]$ denote its left and right $x$-ordinates, $y_{i}$ its
height, i.e., $y_{i}=\varphi(x_{i})\vee\varphi(x_{i+1})$,
$\underline{y}_{i}$ the height of the smaller of its two immediate
neighbours, i.e.,
$\underline{y}_{i}=\varphi(x_{i})\wedge\varphi(x_{i+1})$, and let
$d_{i}=x_{i+1}-x_{i}$, $\delta_{i}=d_{i}y_{i}/\underline{y}_{i}$.
(Symbols $\wedge$ and $\vee$ means `min' and max' throughout the
paper.) All these numbers are computed beforehand and defined as
constants in the program. Note that the region labelled $-N-1$ (resp. $N$) is
the left tail (resp. right tail) truncated at $x=x_{-N}$ (resp. at
$x=x_{N}$).

To sample $X\sim N(0,1)$, one may proceed as follows: choose randomly
region $i$, sample the point $(X,Y)$ uniformly within the chosen
region, and accept $X$ if $Y\leq\varphi(X)$; otherwise
repeat. However, if the chosen region is a rectangle, most of the
computation can be bypassed: one may first simulate $Y$, i.e., draw
$U\sim U[0,1]$ and set $Y=y_{i}U$, without simulating $X$, and check
that the realisation $y$ of $Y$ is such that $y\leq\underline{y}_{i}$;
recall that $\underline{y}_{i}=\varphi(x_{i})\wedge\varphi(x_{i+1})$.
If this condition is fulfilled, then the realised pair $(x,y)$ must be
accepted whatever the value of $x$. Furthermore, one can recycle the
realisation $u$ of $U$, and therefore avoid drawing a second uniform
variate, by simply setting $x=x_{i}+\delta_{i}u$.

In short, with high probability, the algorithm only performs the following
basic operations: 

\begin{algorithm}[H]
\begin{algorithmic}
\STATE draw~a~random~integer~$i$~uniformly~in~range~$-N-1,\ldots, N$
\IF {$i<N$~and~$i>-N-1$}
    \STATE sample~$u\sim U[0,1]$
    \STATE $y\gets y_{i}*u$
    \IF {$y\leq\underline{y}_{i}$}
        \RETURN $x_{i}+\delta_{i}*u$
    \ENDIF
\ENDIF 
\end{algorithmic}
\end{algorithm}

A complete outline of the algorithm is given in Appendix A. When the
condition $y\leq\underline{y}_{i}$ is not fulfilled, one must sample
$X$ and check that $(X,Y)$ is indeed under the curve of the $N(0,1)$
density. Similarly, when the chosen region is either the left or right
tail, one may use Devroye's algorithm in order to simulate $X$.

Note that this algorithm has a slightly higher numerical precision
than the rejection algorithms mentioned in the previous section: when
calculating $x=x_i+\delta_i*u$, the absolute error equals the
precision of the random generator that produces $u$, say
$2^{-32}\approx 2.3\times 10^{-10} $ for a 32 bit generator, times the
small number $\delta_i$, which is typically of order $10^{-3}$.

Many algorithms proposed in the literature already use
histograms to construct a good proposal density; see e.g
\citet{Marsaglia:ziggurat}, \citet{ahrens:domains},
\citet{zaman1991grn} or the survey in \citet[Chap. 5]{hormann:book}.
In particular, the above algorithm is similar to the Ziggurat
algorithm \citep{Marsaglia:ziggurat,Marsaglia:zig2000}, which is the
default Gaussian sampler in much mathematical software, e.g. Matlab or
the GNU Scientific Library, and most similar to
\citet{ahrens1995one}'s algorithm. Both algorithms already use the
idea of using rectangles of equal areas, but the Ziggurat algorithm is
based on horizontal rectangles, while \citet{ahrens1995one}'s
algorithm is based on vertical ones, as above. This seemingly
innocuous variation greatly facilitates the extension to truncated
densities, as explained in next section.

\subsection{Extension to truncated Gaussians\label{sub:Ext-tg}}

For a fixed truncation point $a$, let $l_a$ denote the index of the
region that contains $a$. To adapt the above algorithm to the
truncated density (\ref{eq:tn}), one may choose an integer $i_a$ such
that $i_{a}\leq l_a$, sample randomly one region among
$i_{a},i_{a}+1,\ldots,N$, and proceed as explained above. In addition,
for the regions $i_a,\ldots,l_a$, one must reject the random point
$(X,Y)$ if $X<a$, as described in Appendix A.

The difficulty is to define $i_{a}$ in such a way that (a) the
computation of $i_a$ is quick, and (b) $i_a$ is as close as possible
to $l_a$, so that the overall acceptance rate is as high as possible.  We
propose the following method. We choose a small width $h>0$, and store
beforehand in an integer array the following quantities:

\[
j_{k}=\max\left\{ i:\, x_{i}\leq kh\right\} ,\qquad\mbox{for all
}k\mbox{ such that }kh\in[a_{\min},a_{\max}]
\]
for some interval $[a_{\min},a_{\max}]$. As said before, all these
constants are computed separately, and hard-coded in the program.
Then, provided $a\in[a_{\min},a_{\max}]$, $i_{a}$ is computed
 as\[
i_{a}=j_{\left\lfloor a/h\right\rfloor }\]
 where $\left\lfloor \cdot\right\rfloor $ stands for  the floor
 function. Provided $h\leq \min_i(x_{i+1}-x_i)$, each interval $[kh,(k+1)h)$
contains at most one $x_i$, so that either $i_a=l_a$ or
$i_a=l_a-1$. This means that, when choosing randomly between
regions $i_a,i_a+1,\ldots,N$, one must treat separately
the two leftmost regions $i_a$ and $i_a+1$, and perform the 
additional check mentioned above, i.e., $x\geq a$,  but the other regions
$i_a+2,\ldots,N$ can be treated exactly as explained in the 
previous section. 

In our simulations, we set $a_{\min}=-2$, $a_{\max}=x_{N-20}$, and
$h=x_1-x_0$, that is, the smallest of the interval ranges
$(x_{i+1}-x_i)$.  A complete outline of the algorithm is given in
Appendix A.

\subsection{Results}

\begin{figure}
\begin{centering}
\includegraphics[scale=0.5]{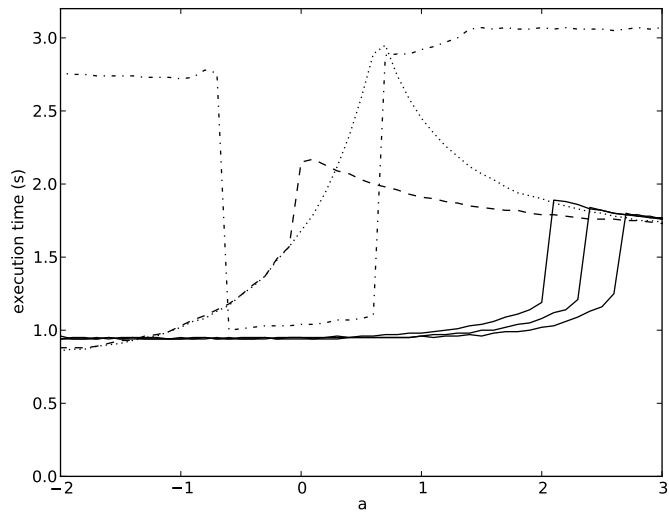}
\par\end{centering}

\caption{\label{fig:Execution-time}Execution time (seconds) vs
  truncation point, for $10^{8}$ simulations of Devroye's algorithm
  (dotted line), Geweke-Robert sampler (dashed line), the inverse
  transform algorithm (based on the 
  inverse transform approximation of \cite{wichura1988algorithm}, as
  implemented in the GSL library, dash-dotted line), and our algorithm
  (solid lines), with, from left to right, $N_{s}=1000$, $N_{s}=2000$,
  $N_{s}=4000$.}

\end{figure}

We implemented our algorithm, the inverse transform algorithm,
Devroye's algorithm (using cut-off value $a_{0}=0.65$) and
Geweke-Robert's algorithm in $ $C, using the GNU Scientific
library (GSL). Figure \ref{fig:Execution-time}
plots the execution time of $10^{8}$ runs on a 2.8 Ghz desktop
computer, for each algorithm and for different values of the
truncation point $a$. Our algorithm appears to be up to two times
faster than Geweke-Robert's algorithm, and up to three times faster
than Devroye's algorithm and the inverse transform method. The three
solid lines correspond to different sizes $N_{s}$,
$N_{s}=1000,\,2000,\,4000$, from left to right, of the five arrays
containing the constants $x_{i}$, $y_{i}$, $\underline{y}_{i}$,
$d_{i}$, $\delta_{i}$; note that only those values such that
$x_{i},x_{i+1}\in[a_{\min},a_{\max}]$ need to be stored, hence
$N_{s}<2N$. Figure \ref{fig:Execution-time} shows that increasing
$N_{s}$ only improves the execution time for a tiny interval of $a$
values, so there may be little point in increasing $N_{s}$ further
than, say, $4000$. For $N_{s}=4000$, the acceptance rate is typically
above $0.99$ or even $0.999$ for most values of
$a\in[a_{\min},a_{\max}]$.  The increase of the computational cost for
$a>2$ is due to the increasing probability of sampling $X$ from the
right tail using Devroye's algorithm. Outside of the range of the
represented interval, the rejection algorithms have a similar
computational cost, as they perform the same operations. For
$N_{s}=4000$ and a double precision implementation, the total memory
cost of the algorithm is 162 kB, a small fraction of the memory cache
of most modern CPU's. (A CPU memory cache is a small, fast memory
where a CPU stores data used repetitively.)

\subsection{Truncation to a finite interval $[a,b]$\label{sub:finite}}

The extension to finite truncation intervals $[a,b]$ is
straightforward.  First, one determines a region index $i_{a}$
(resp. $i'_{b}$) such that either region $i_a$ or region $i_a+1$
(resp. region $i'_b$ or region $i'_b-1$) contains $a$ (resp. $b$),
using the table look-up method described in \ref{sub:Ext-tg}. Then,
one proceeds as above, choosing randomly a region in the range
$i_a,\ldots,i'_b$, and so on.

However, a difficulty arises if $(b-a)$ is small. Suppose for instance
that $a$ and $b$ fall in the same region, and that $b-a$
is small with respect to the width of the region. Then, if one samples
uniformly point $(X,Y)$ within that region, the probability that $a\leq
X\leq b$ may be arbitrarily small. 
% More generally, if $i_{b}'-i_{a}=k$,
% the probability that the proposed values falls outside $[a,b]$ should
% be, at worst, approximately $2/(k+1)$.

We propose the following work-around: when $i_{b}'-i_{a}\leq k_{\min}$,
say $k_{\min}=5$, use instead a rejection algorithm based on \citet{Devroye:book}'s
exponential proposal, but truncated to $[a,b]$, i.e., \begin{equation}
q(x)=\frac{\lambda\exp(-\lambda x)}{\exp(-\lambda a)-\exp(-\lambda b)}I(a\leq x\leq b).\label{eq:truncexp}\end{equation}
with $\lambda=a$ (resp. $\lambda=b$) when $b>0$ (resp. when $b\leq0$).
The advantage of this approach is that it gives an acceptance rate
close to $1$ whatever the values of $a$, and $b$, subject to $i_{b}'-i_{a}\leq k_{\min}$;
i.e., whether $a$ and $b$ are both in the same tail, or both close
to $0$. In the latter case, $q$ should be close numerically to
a uniform distribution.

\section{Bi-dimensional case: semi-finite intervals\label{sec:bitgsemi}}

We now consider the simulation of $X=(X_{1},X_{2})\sim N_{2}(\mu,\Sigma)$,
subject to $X_{1}\geq a_{1}$ and $X_{2}\geq a_{2}$. Without loss
of generality, we set $\mu=(0,0)'$,
 \[
\Sigma=\left(\begin{array}{cc}
1 & \rho\\
\rho & 1\end{array}\right),\]
and assume that $a_{1}\geq a_{2}$; if necessary, swap components
to impose the last condition. The joint density of the considered
truncated density is, up to a constant:
 \begin{equation}
p(x_{1},x_{2})\propto\exp\left\{ -\frac{1}{2\nu^{2}}\left(x_{1}^{2}+x_{2}^{2}-2\rho x_{1}x_{2}\right)\right\} I\left(x_{1}\geq a_{1};x_{2}\geq a_{2}\right),\label{eq:bitg}\end{equation}
where the short-hand $\nu^{2}=1-\rho^{2}$ will be used throughout
the rest of the paper. The conditional distribution of $X_{2}|X_{1}=x_{1}$
is a univariate Gaussian $N(\rho x_{1},\nu^{2})$ truncated to $X_{2}\geq a_{2}$,
which we denote from now on $TN_{[a_{2},\infty)}(\rho x_{1},\nu^{2})$.
A common misconception is that the marginal density of $X_{1}$ is
also a truncated Gaussian density, although standard calculus leads
to: 
\begin{equation}
p(x_{1})\propto\varphi(x_{1})\Phi\left(\frac{\rho x_{1}-a_{2}}{\nu}\right)I(x_{1}\geq a_{1}).\label{eq:marg}\end{equation}

In order to simulate from (\ref{eq:bitg}), a natural strategy is
to derive a rejection algorithm for the marginal (\ref{eq:marg}),
and to simulate $X_{2}$ conditional on $X_{1}$, using the algorithm
we developed in Section 2. This is basically the approach adopted
here, although we shall see that, in some cases, it is preferable
to derive a rejection sampler for the joint distribution (\ref{eq:bitg}).
We mention briefly that universal bivariate samplers exist, see e.g.
\citet{Hormann:autobi} or \citet{leydold2000asr}, but
as in the univariate case our objective is to design a specialised
algorithm that runs faster (i.e., does not require a set-up time),
for situations where only one random vector must be generated. 

To derive a proposal distribution for (\ref{eq:marg}), we substitute
the $\Phi(\cdot)$ factor with a simpler expression derived from the
two following straightforward inequalities:
 \begin{equation}
  \frac{1}{2}\leq\Phi(x)\leq1\quad\mbox{for
  }x\geq0\label{eq:inp},
\end{equation}
\begin{equation}
  \Phi(x)\leq c(x_{0})\varphi(x)\quad\mbox{for }x\leq
  x_{0}\leq0,\label{eq:inm}
\end{equation}
where $c(x_{0})=(\sqrt{\pi/2})\wedge(-1/x_{0})$, for $x_{0}<0$,
$c(0)=\sqrt{\pi/2}$. We now distinguish between cases where the argument
of $\Phi(\cdot)$ in (\ref{eq:marg}) is positive, negative, or both,
over the range of possible values for $x_{1}.$ We consider the following
cases, and treat them separately:
\begin{itemize}
\item case $S^{+}$: either $\rho\geq0$ and $\rho a_{1}-a_{2}\geq0$, or
$\rho<0$ and $a_{1}\leq\Phi^{-1}(1/3)\approx-0.4307$ 
\item case $S^{-}$: $\rho<0$, $\rho a_{1}-a_{2}\leq0$, and $a_{1}>\Phi^{-1}(1/3)\approx-0.4307$.
\item case $M^{+}:$ $\rho\geq0$ and $\rho a_{1}-a_{2}<0$.
\item case $M^{-}$: $\rho<0$, $\rho a_{1}-a_{2}>0$, and
  $a_{1}>\Phi^{-1}(-1/3)\approx -0.4307$.
\end{itemize}
where  `S' stands for `Simple', and `M' for `Mixture', as we elaborate
below. 

We now prove that, in each case, it is possible to derive a rejection
algorithm with an acceptance rate bounded from below for all values of 
 $\rho$, $a_{1}$ and $a_{2}$.

\subsection{Case $S^{+}$}

Assuming first $\rho\geq0$ and $\rho a_{1}-a_{2}\geq0$, then, according
to (\ref{eq:inp}),\begin{equation}
\Phi\left(\frac{\rho x_{1}-a_{2}}{\nu}\right)\in[1/2,1]\label{eq:phiin}\end{equation}
 for all $x\geq a_{1},$ which suggests the following proposal distribution:
\[
q_{S^{+}}(x_{1})\propto\varphi(x_{1})I(x_{1}\geq a_{1}),\]
i.e., a $TN_{[a_{1},+\infty)}(0,1)$ distribution, in order to sample
from the marginal $p(x_{1})$. For a given $x_{1}$ simulated from
$q_{S^{+}}$, the acceptance probability equals (\ref{eq:phiin}),
hence the acceptance rate of such a rejection algorithm equals
 \begin{equation}
\int_{a_{1}}^{+\infty}\Phi\left(\frac{\rho x_{1}-a_{2}}{\nu}\right)\frac{\varphi(x_{1})}{\Phi(-a_{1})}\, dx_{1},\label{eq:ar}\end{equation}
 and is larger than $1/2$ by construction. 

However, it is more efficient to simulate jointly $(X_{1},X_{2})$
as follows: sample $X_{1}\sim TN_{[a_{1},\infty)}(0,1)$, $X_{2}|X_{1}=x_{1}\sim N(\rho x_{1},\nu^{2})$,
and accept if $X_{2}\geq a_{2}$; otherwise repeat. It is easy to
check that the latter rejection algorithm has exactly the same acceptance
rate, i.e., (\ref{eq:ar}), as the former, but it is faster, because
it does not perform any evaluation of $\Phi$, and because $X_{2}$
is obtained `for free', i.e., a second step is not required to generate
$X_{2}$. 

When $\rho<0$, the argument of $\Phi$ in (\ref{eq:ar}) is not positive
for all $x_{1}\geq a_{1}$, but the integral is still larger than
$1/2$ provided $a_{1}\leq\Phi^{-1}(1/3)$. To establish this property,
one may remark that (\ref{eq:ar}) is the probability that $X_{2}\geq a_{2}$,
conditional on $X_{1}\geq a_1$, provided $(X_{1},X_{2})\sim N_{2}(\mu,\Sigma)$.
This probability decreases with respect to $a_{2}$, $a_{2}\leq a_{1}$,
and, for $a_{2}=a_{1}$, this probability decreases with respect to
$-\rho$ and $a_{1}$. Finally, for $\rho=-1$ and $ $$a_{1}=a_{2}=\Phi^{-1}(1/3)$,
this probability equals $1/2.$ Thus, we use
 Algorithm $S^{+}$ also when $\rho<0$ and $a_{1}\leq\Phi^{-1}(1/3)$.

\subsection{Case $S^{-}$\label{sub:Case-Sm} }

If $\rho<0$ and $\rho a_{1}-a_{2}\leq0$, then inequality
(\ref{eq:inm}) holds for all values of the argument $ $$x=(\rho
x_{1}-a_{2})/\nu$, and for $x_{0}=(\rho a_{1}-a_{2})/\nu$. This
suggests the following algorithm: sample $X_{1}$ from proposal
density
\begin{eqnarray*}
  q_{S^{-}}(x_{1}) & \propto & \varphi(x_{1})\varphi\left(\frac{\rho x_{1}-a_{2}}{\nu}\right)I(x_{1}\geq a_1)\\
  & \propto & \varphi\left(x_{1};\rho a_{2},\nu^{2}\right)I(x_{1}\geq
  a_1)\end{eqnarray*} 
that is, the density of truncated Gaussian
distribution $TN_{[a_{1},\infty)}(\rho a_{2},\nu^{2})$, and accept
with probability: \begin{equation} \psi\left(-\frac{\rho
      x_{1}-a_{2}}{\nu}\right)/c\left(\frac{\rho
      a_{1}-a_{2}}{\nu}\right),\label{eq:probsm}\end{equation} where
$\psi(x)=\Phi(-x)/\varphi(x)$. The acceptance rate is
then:\begin{equation} E_{TN_{[a_{1},\infty)}(\rho
    a_{2},\nu^{2})}\left[\psi\left(-\frac{\rho
        X_{1}-a_{2}}{\nu}\right)\right]/c\left(\frac{\rho
      a_{1}-a_{2}}{\nu}\right).\label{eq:arsm}\end{equation} We show
formally in Appendix B1 that this acceptance rate admits a lower bound
which is larger than $0.416$; our numerical studies indicate that the
optimal lower bound may be $1/2$.  Intuitively, the idea behind
inequality (\ref{eq:inm}) is that $\Phi(-x)\approx\varphi(x)/x$ for
large values of $x$, hence the true marginal density $p(x_1)$ behaves
like a Gaussian density density times $1/(a_{2}-\rho x_{1})$, but the
latter factor varies slowly with respect to a Gaussian density, so it
can be discarded in the proposal.

\subsection{Case $M^{-}$\label{sub:Case-MM}}

If $\rho<0$ and $\rho a_{1}-a_{2}>0$, the quantity $(\rho x_{1}-a_{2})/\nu$
takes positive and negative values for $x_{1}\geq a_{1}$. To combine
both inequalities, one may use a mixture proposal:
\begin{equation}
q_{M^{-}}(x_{1})\propto\varphi(x_{1})I\left(\rho
  x_{1}-a_{2}>0\right)+\sqrt{\frac{\pi}{2}}\varphi(x_{1})\varphi\left(\frac{\rho
    x_{1}-a_{2}}{\nu}\right)I\left(\rho
  x_{1}-a_{2}<0\right)\label{eq:mixm}
\end{equation}
subject to $x_{1}\geq a_{1}$. To sample from the mixture proposal,
choose component $1$ (corresponding to the first term above), with
probability $\omega_{1}/(\omega_{1}+\omega_{2})$, with \[
\omega_{1}=\Phi(a_{2}/\rho)-\Phi(a_{1})\]
\[
\omega_{2}=\frac{\nu}{2}\exp\left\{ -\frac{a_{2}^{2}}{2}\right\} \Phi\left(-\frac{a_{2}\nu}{\rho}\right)\]
and choose component 2 otherwise. If component 1 is chosen, one can
use the same shortcut as in Algorithm $S^{+}$, that is, draw $X_{1}\sim TN_{[a_{1},a_{2}/\rho]}(0,1)$
and $X_{2}|X_{1}=x_{1}\sim N(\rho x_{1},\nu^{2})$, and accept the
simulated pair $(x_{1},x_{2})$ if $x_{2}\geq a_{2}$. If component
2 is chosen, the proposed value for $X_{1}$ is drawn from a $TN_{[a_{2}/\rho,\infty)}(\rho a_{2},\nu^{2})$
distribution, and is accepted with probability \[
\sqrt{\frac{2}{\pi}}\psi\left(-\frac{\rho x_{1}-a_{2}}{\nu}\right).\]
When a draw $x_{1}$ is accepted, it is completed with $x_{2}$ drawn
from $X_{2}|X_{1}=x_{1}\sim TN_{[a_{2},+\infty)}(\rho x_{1},\nu^{2})$.

The acceptance rate of this algorithm is a weighted average (with
weights given by $\omega_{1}$ and $\omega_{2}$) of the acceptance
rate of Component 1, which is larger than $1/2$ by construction,
and the acceptance rate of algorithm $S^{-}$ for $a_{2}=\rho a_{1}$,
which is also bounded from below, as explained in the previous
subsection.

\subsection{Case $M^{+}$\label{sub:Case-Mp}}

If $\rho\geq0$ and $\rho a_{1}-a_{2}\leq0$, then again $(\rho x_{1}-a_{2})/\nu$
take both negative and positive values for $x_{1}\geq a_{1}$, which
suggests that we use a mixture proposal similar to (\ref{eq:mixm}). Unfortunately,
the acceptance rate may be arbitrarily small in that case. Exact calculations
are omitted for the sake of space, but it can be shown that the mode
of $p(x_{1})$ can be arbitrary far from $x_{1}=a_{2}/\rho$, the
point where the two components intersect, which gives an arbitrary
small acceptance rate. 

Instead, we substitute (\ref{eq:inm}) with a slightly different
inequality:
\begin{equation}
\Phi(x)\leq d(x_{0})\varphi(x)e^{\lambda x}\quad\mbox{for }x_{0}\leq
x\leq0\label{eq:inmp}
\end{equation}
where $\lambda=0.68$, $d(x_{0})=(\sqrt{\pi/2})\vee\chi(-x_{0})$,
and $\chi(x)=e^{\lambda x}\Phi(-x)/\varphi(x)$. This inequality stems
from straightforward calculus. Other values of $\lambda$ are also
valid, but in our numerical experiments, $\lambda=0.68$ seemed to
be close to optimal, in terms of minimum acceptance rate. 

This inequality leads to the following proposal mixture density: 
\begin{eqnarray*}
q_{M^{+}}(x_{1}) & \propto & \varphi(x_{1})I\left(\rho x_{1}-a_{2}\geq0\right)\\
 &  & +\varphi(x_{1})\varphi\left(\frac{\rho x_{1}-a_{2}}{\nu}\right)\exp\left(\frac{\lambda(\rho x_{1}-a_{2})}{\nu}\right)d\left(\frac{\rho a_{1}-a_{2}}{\nu}\right)I\left(\rho x_{1}-a_{2}<0\right)\end{eqnarray*}
subject to $x_{1}\geq a_{1}$. The second term is proportional to
a $N(\theta,\nu^{2})$ density, with $\theta=\rho(a_{2}+\lambda\nu)$.
To sample from this mixture, choose component $1$, with probability
$\tau_{1}/(\tau_{1}+\tau_{2})$, choose component 2 otherwise, where
\[
\tau_{1}=\Phi(-a_{2}/\rho)\]
\begin{eqnarray*}
\tau_{2} & = & \frac{\nu}{\sqrt{2\pi}}\left\{ \Phi\left(\frac{a_{2}/\rho-\theta}{\nu}\right)-\Phi\left(\frac{a_{1}-\theta}{\nu}\right)\right\} \\
 &  & \qquad \exp\left\{ \frac{\theta^{2}-a_{2}^{2}-2\lambda\nu a_{2}}{2\nu^{2}}\right\} d\left(\frac{\rho a_{1}-a_{2}}{\nu}\right).\end{eqnarray*}
If component 1 is selected, draw $X_{1}\sim TN_{[a_{2}/\rho,+\infty)}(0,1)$,
$X_{2}|X_{1}=x_{1}\sim N(\rho x_{1},\nu^{2})$, and accept simulated
pair $(x_{1},x_{2})$ if $x_{2}\geq a_{2}$. Otherwise, draw $X_{1}\sim TN_{[a_{1},a_{2}/\rho]}(\theta,\nu^{2})$,
and accept with probability \[
\chi\left(\frac{a_{2}-\rho x_{1}}{\nu}\right)/d\left(\frac{a_{2}-\rho a_{1}}{\nu}\right),\]
and, upon acceptance, complete with \[
X_{2}|X_{1}=x_{1}\sim TN_{[a_{2},+\infty)}(\rho x_{1},\nu^{2}).\]
We show formally in Appendix B2 that the acceptance rate of this algorithm
is bounded from below by $0.22$, and we found numerically that the
optimal lower bound seems to be $1/2$, see Section \ref{sub:Results}.

\subsection{Computational cost}

The above algorithms, except algorithm $S^{+}$, involve a few evaluations
of function $\Phi$, which is expensive. But such evaluations can
be bypassed in most cases. In algorithm $S^{-}$ for instance, given
the expression of acceptance probability (\ref{eq:probsm}), one should
accept the proposed value $x_{1}$ if and only if 

\[
\Phi\left(rx_{1}+s\right)\geq ut(x_{1})\]
where $u$ is an uniform variate, and the exact expression of $r,$
$s$, and $t$ are easily deduced from (\ref{eq:probsm}). If good,
fast approximations of $\Phi$ are available, such that $\underline{\Phi}(x)\leq\Phi(x)\leq\overline{\Phi}(x)$,
it is enough to check that $\underline{\Phi}\left(rx_{1}+s\right)\geq ut(x_{1})$
(resp. $\overline{\Phi}\left(rx_{1}+s\right)< ut(x_{1})$) to accept
(resp. reject) $x_{1}$. It is only when $ut(x_{1})$ is very close
to $\Phi\left(rx_{1}+s\right)$ that an exact evaluation of $\Phi\left(rx_{1}+s\right)$
is required. 

Such fast, good approximations of $\Phi$ may be deduced from 
the tables of our univariate algorithm, see Section \ref{sub:Ext-tg}.
Specifically, and using the same notations as Section \ref{sub:Ext-tg},
let $z=rx_{1}+s$, and assume that $z\in[a_{\min},a_{\max}]$, then
one may set $\overline{\Phi}(z)=A(j_{\left\lfloor z/h\right\rfloor
}+1)$,
where $A(i)>\Phi(x_{i+1})$ denotes the total area of all the regions up to region
$i$, and may be computed beforehand and hard-coded in the program
like the other constants  $y_i$, $d_i$ and so on. One may define similarly 
$\underline{\Phi}(z)$ using $\Phi(z)=1-\Phi(-z)$. When $z\notin[a_{\max},a_{\min}]$,
one may use  the expansion of \citet[p. 932]{handbook_math_functions}
to derive the following upper and lower approximations \[
\overline{\Phi}(z)=-\frac{\varphi(z)}{z}\left\{ 1-\frac{1}{z^{2}}+\frac{3}{z^{4}}\right\} ,\qquad\underline{\Phi}(z)=-\frac{\varphi(z)}{z}\left\{ 1-\frac{1}{z^{2}}+\frac{3}{z^{4}}-\frac{15}{z^{6}}\right\} ,\]
for $z<0$. (For $z>0$, similar formulae are obtained using $\Phi(z)=1-\Phi(-z$)).
Note also that one does not necessarily have to compute all the
terms of these expressions. For instance, if $\overline{\Phi}_{1}(\alpha x_{1}+\beta)<u\upsilon(x_{1})$,
where $\overline{\Phi}_{1}(z)=-\varphi(z)/z$, then the proposed value
can be rejected without computing the remaining terms. This principle
can be used for each term of the expansion, but, on the other hand,
it is preferable not to expand further the expressions above, since
they work well already for reasonable values of $a_{\min}$ and $a_{\max}$,
and since that would define diverging series. 

Provided the above strategy is implemented, the algorithms proposed
in the section are reasonably fast, since they only involve a few
basic operations, and their acceptance rate is greater than $1/2$
for all parameters. Algorithms $M^{+}$ and $M^{-}$ are slightly
more expensive, as they require sampling a mixture index, but note
that the same strategy can be implemented in order to avoid with good
probability the evaluation of functions $\Phi$ appearing in the expression
of the mixture weights.

\subsection{Numerical illustration\label{sub:Results}}

\begin{figure}

\centering{}\includegraphics[scale=0.5]{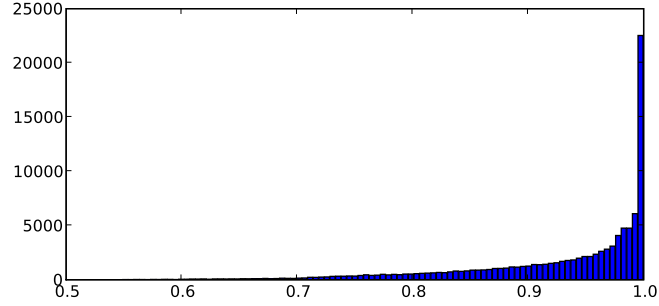}
\caption{\label{fig:histar}Histogram of acceptance rates corresponding to
$10^{5}$ simulated vectors $(a_{1},a_{2},\rho)$, where $\rho\sim U[0,1]$,
$a_{1}$, $a_{2}\sim N(0,1)$, conditional on $a_{1}\geq a_{2}$. }

\end{figure}

We simulated $10^{5}$ parameters $(a_{1},a_{2},\rho)$, where $\rho\sim U[-1,1]$,
$a_{1}$, $a_{2}\sim N(0,s^{2})$, conditional on $a_{1}\geq a_{2}$.
For each parameter, we evaluated the acceptance rate of our algorithm
by computing the average acceptance probability over the proposed
draws generated by $1000$ runs. Figure \ref{fig:histar} reports
the histogram of the acceptance rates for $s=1$; larger values of
$s$ give histograms that are even more concentrated around $1$.

In this simulation exercise, $90\%$ of the acceptance rates are above
$0.8$ and $99\%$ are above $0.65$; none is lower than $1/2$.

\section{Bi-dimensional case: finite intervals}

We now consider the simulation of $X=(X_{1},X_{2})\sim N_{2}(0,\Sigma)$,
with \[
\Sigma=\left(\begin{array}{cc}
1 & \rho\\
\rho & 1\end{array}\right),\]
conditional on $X_{1}\in[a_{1},b_{1}]$ and $X_{2}\in[a_{2},b_{2}]$,
where, without loss of generality, $\rho\geq 0$. 
As in the previous section, the main difficulty is to sample from
the marginal density of $X_{1}$:
\begin{equation}
p(x_{1})\propto\varphi(x_{1})\left[\Phi(\alpha
  x_{1}+\beta_{1})-\Phi(\alpha x_{1}+\beta_{0})\right]
I(a_1\leq x_1 \leq b_1)
\label{eq:margab}\end{equation}
where $\alpha=\rho/\nu\geq 0$, $\beta_{1}=-a_{2}/\nu$, $\beta_{0}=-b_{2}/\nu$
and $\beta_{1}\geq\beta_{0}$, since the distribution of $X_{2}$
conditional on $X_{1}=x_{1}$ is the univariate truncated Gaussian
distribution $TN_{[a_{2},b_{2}]}(\rho x_{1},\nu^{2})$.

We shall consider two situations, according to whether $\beta_{1}-\beta_{0}<\Delta$
or not; we take $\Delta=2$, which seems to close to the optimal value
in our simulations, in terms of minimum acceptance rate. 

\begin{figure}
\begin{centering}
\includegraphics[scale=0.5]{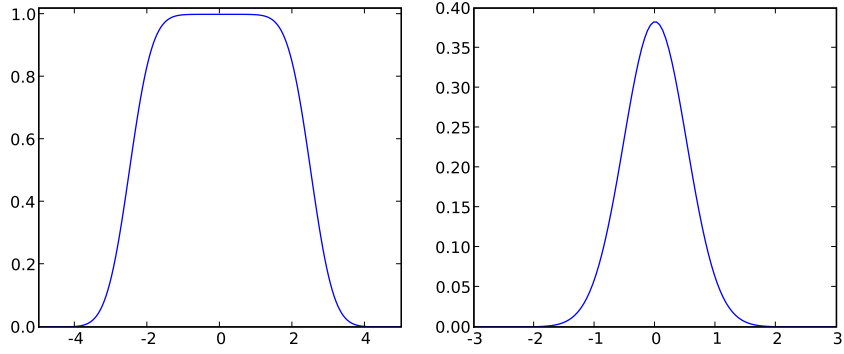}
\par\end{centering}

\caption{\label{fig:kappa}Function $\kappa(x_{1})=\Phi(\alpha x_{1}+\beta_{1})-\Phi(\alpha x_{1}+\beta_{0})$
, for $(\alpha,\beta_{0},\beta_{1})=(2,-5,5)$ (left), and $ $$(\alpha,\beta_{0},\beta_{1})=(2,-0.5,0.5)$
(right)}

\end{figure}

When $\beta_{1}-\beta_{0}$ is small (case $T$), one may Taylor expand
the second factor of (\ref{eq:margab}), which is denoted $\kappa$
from now on, into: \[
\kappa(x_{1})=\Phi(\alpha x_{1}+\beta_{1})-\Phi(\alpha x_{1}+\beta_{0})\approx(\beta_{1}-\beta_{0})\varphi\left(\alpha x_{1}+\frac{\beta_{0}+\beta_{1}}{2}\right),\]
hence $\kappa$ behaves like a Gaussian density, see the right panel of
Figure \ref{fig:kappa}. Therefore, $p(x_{1})$ is also well approximated
by a Gaussian, which is the basis of algorithm T detailed in 4.2.

Otherwise, when $\beta_{1}-\beta_{0}$ is large, $\kappa$ behaves
like the curve plotted in the left panel of Figure \ref{fig:kappa}. In
this case, called $M^{3}$ below, we cut $\kappa$ into at most three
pieces, and derive a mixture proposal, using ideas similar to the
previous section. 

\subsection{Case $M^{3}$}

Let $\gamma_{i}=-\beta_{i}/\alpha$, for $i=0,1$, and $\upsilon=(\beta_{1}-\beta_{0})/2$;
note $\gamma_{1}\leq\gamma_{0}$ since $\beta_{0}\leq\beta_{1}$.
We may divide the curve of $\kappa$ into three parts, so as to re-use
the same ideas as in Section \ref{sec:bitgsemi}, that is, deriving
a mixture of Gaussian proposals. Specifically, 
\begin{itemize}
\item for $x_{1}<\gamma_{1}$, one shows easily that
 \[
\left\{ 2\Phi(\Delta)-1\right\} \Phi(\alpha x_{1}+\beta_{1})\leq\kappa(x_{1})\leq\Phi(\alpha x_{1}+\beta_{1}).\]
We know already that target density $\varphi(x_{1})\Phi(\alpha x_{1}+\beta_{1})I(x_{1}<\gamma_{1})$
can be simulated efficiently using a particular Gaussian proposal
density, see Component 2 of Algorithm $M^{+}$ in Section \ref{sub:Case-Mp}.
The above inequality indicates that, using the same proposal, one
should obtain an acceptance rate which is at least $2\Phi(\Delta)-1$
times the minimum acceptance rate of $M^{+}$, that is, $\Phi(\Delta)-1/2\approx0.477.$
\item for $x_{1}\in[\gamma_{1},\gamma_{0}]$, $\kappa$ is roughly flat,
and \[
\Phi(\Delta)-1/2\leq\kappa(x_{1})\leq2\Phi(\upsilon)-1\leq1.\]
This suggests using proposal distribution $X_{1}\sim TN_{[\gamma_{1}\vee a_{1},\gamma_{0}\wedge b_{1}]}(0,1)$,
and accept realisation $x_{1}$ with probability $\kappa(x_{1})/\left\{ 2\Phi(\upsilon)-1\right\} $.
The acceptance probability is then bounded from below by $\Phi(\Delta)-1/2\approx0.477$. 
\item for $x_{1}>\gamma_{0}$, one has: \[
\left\{ 2\Phi(\Delta)-1\right\} \Phi(-\alpha x_{1}-\beta_{0})\leq\kappa(x_{1})\leq\Phi(-\alpha x_{1}-\beta_{0}).\]
Again, one may use the same proposal as for Component 2 of algorithm
$M^{-}$ , see \ref{sub:Case-MM}, which should lead to an acceptance
rate which is larger than $\Phi(\Delta)-1/2\approx0.477$. 
\end{itemize}
The principle of Algorithm $M^{3}$ is therefore to draw from a mixture
of at most three components, the relative weights of which are given
below, and given the chosen component, to use one of the three strategies
described above. 

Denote $\zeta_{l}$, $\zeta_{c}$, $\zeta_{r}$, the unnormalised
weights of the left, centre, and right components, respectively. In case
$b_{1}\vee\gamma_{1}>0$, one has
 \[
\zeta_{l}=\frac{\nu\, d(\alpha a_{1}+\beta_{1})}{\sqrt{2\pi}}\exp\left(\frac{m_{l}^{2}-a_{2}^{2}-2\lambda\nu a_{2}}{2\nu^{2}}\right)\left\{ \Phi\left(\frac{\gamma_{1}-m_{l}}{\nu}\right)-\Phi\left(\frac{a_{1}-m_{l}}{\nu}\right)\right\} I(\gamma_{1}\geq a_1)\]
with $m_{l}=\rho(a_{2}+\lambda\nu)$ and function $d$ was defined
in Section \ref{sub:Case-Mp}; Otherwise, one obtains the same expression,
but with $\lambda$ set to $0$, i.e., \[
\zeta_{l}=\frac{\nu}{2}\exp\left(-\frac{a_{2}^{2}}{2}\right)\left\{ \Phi\left(\frac{\gamma_{1}-m_{l}}{\nu}\right)-\Phi\left(\frac{a_{1}-m_{l}}{\nu}\right)\right\} I(\gamma_{1}\geq a_1)\]
with $m_{l}=\rho a_{2}$. In all cases, 

\[
\zeta_{c}=\left\{ 2\Phi(\upsilon)-1\right\} \left\{ \Phi(\gamma_{0}\wedge b_1)-\Phi(\gamma_{1}\vee a_1)\right\} I(b_{1}>\gamma_{1};a_{1}<\gamma_{0}),\]
\[
\zeta_{r}=\frac{\nu}{2}\exp\left(-\frac{b_{2}^{2}}{2}\right)\left\{ \Phi\left(\frac{b_{1}-\rho b_{2}}{\nu}\right)-\Phi\left(\frac{\gamma_{0}-\rho b_{2}}{\nu}\right)\right\} I(b_{1}>\gamma_{0}).\]

One may show that the acceptance rate of this algorithm is bounded
from below by $1/2-\Phi(-\Delta)\approx0.477$; simulations suggest
this bound is optimal. We omit the exact calculations, as they are
similar to those of previous algorithms. We managed to obtain this
result under the following assumptions: (i) $\rho\geq0$; (ii) $b_{2}\geq0$
and (iii) either $a_{2}\geq a_{1}$ or $b_{1}\leq0$. It is always
possible to enforce such conditions, by either swapping $X_{1}$ and
$X_{2}$, or changing their signs, or both. We note also that, in
most of our simulated exercises, at least one component of this mixture
is empty, and often two of them are, which makes it possible to skip
calculating the weights and simulating the mixture index.

\subsection{Case $T$\label{sub:Case-T}}

As explained above, when $\beta_{1}-\beta_{0}<\Delta$, a good Gaussian
approximation of $p(x_{1})$ is \[
q(x_{1})\propto\varphi(x_{1})\varphi\left(\alpha x_{1}+\frac{\beta_{0}+\beta_{1}}{2}\right)\]
 that is, a $N(m,s^{2})$ density with \[
(m,s^{2})=\left(-\frac{\alpha(\beta_{0}+\beta_{1})}{2(1+\alpha^{2})},\frac{1}{1+\alpha^{2}}\right).\]
In our experiments, this approximation appears to be accurate for
all values of $\alpha$, $\beta_{0}$, $\beta_{1}$ (in the sense
that the rejection rate of the algorithm we now describe is always
larger than 0.47 in our simulations, see next subsection). On the
other hand, it seems difficult to apply approximations similar to
those we used before. Instead, we note that $p(x_{1})$ is a log-concave
density \citep{prekopa1973lcm}. This suggests using either exponential
or piecewise exponential proposals, and working out a simplified version
of ARS \citep[Adaptive Rejection Sampling, ][]{GilksWild}.

Specifically, let $\xi$ denote the marginal log-density of $X_{1}$:
\[
\xi(x_{1})=\log\varphi(x_{1})+\log\kappa(x_{1})\]
the derivative of which is easy to compute:\[
\xi'(x_{1})=-x_{1}+\alpha\frac{\varphi(\alpha x_{1}+\beta_{1})-\varphi(\alpha x_{1}+\beta_{0})}{\Phi(\alpha x_{1}+\beta_{1})-\Phi(\alpha x_{1}+\beta_{0})},\]
which leads to the inequality \[
\xi(x_{1})\leq\xi(v)+\xi'(v)(x_{1}-v)\]
for any $x_{1},v\in[a_{1},b_{1}]$. Up to a constant, the right hand
side is the log-density of the truncated Exponential distribution
$\mathrm{Exp}_{[a_1,b_1]}(\lambda)$ defined in (\ref{eq:truncexp}),
with $\lambda=\xi'(v)$; note that $\lambda$ may be negative. Thus,
one may sample $x{}_{1}\sim\mathrm{Exp}_{[a_1,b_1]}\left\{ \xi'(v)\right\} $,
and accept with probability
 \[
\exp\left\{ \xi(x_{1})-\xi(v)-\xi'(v)(x_{1}-v)\right\} .\]
Obviously, the difficulty is to choose $v$. Since the target density
is well approximated by a $TN_{[a_{1},b_{1}]}(m,s^{2})$, and assuming
that $a_{1}\geq m$ (resp. $b_{1}\leq m$), it seems reasonable to
set $v=(m+s)\vee a_{1}$ (resp. $v=(m-s)\wedge b_{1}$). These values
would be optimal if the target density would be equal to its approximation
$TN_{[a_{1},b_{1}]}(m,s^{2})$. In case $a_{1}<m$ and $b_{1}>m$,
i.e., the mean $m$ is within $[a_{1},b_{1}]$, we use instead a mixture
proposal, based on two well chosen points $v,w\in[a_{1},b_{1}]$,
say $v<w$. Thus, \[
\xi(x_{1})\leq\left\{ \xi(v)+\xi'(v)(x_{1}-v)\right\} \wedge\left\{ \xi(w)+\xi'(w)(x_{1}-w)\right\} \]
and one may sample from the density defined as the exponential of
the right hand side above (which is a piecewise exponential distribution),
and accept with probability \[
\exp\left[\xi(x_{1})-\left\{ \xi(v)+\xi'(v)(x_{1}-v)\right\} \wedge\left\{ \xi(w)+\xi'(w)(x_{1}-w)\right\} \right].\]

Note that, in the original ARS algorithm of \citet{GilksWild}, the
proposal is progressively refined by adding a new component each time
the proposed value is rejected. We found however that the acceptance
rate of Algorithm T described above is generally above $1/2$, so
using fixed proposals with at most two components seems reasonable.
Obviously, the good properties of the above algorithm lie in the good
choice of points $v$, $w$, which was made possible by the knowledge
of a good approximation of the target density. 

We were not able to prove formally that the acceptance rate of Algorithm
T is bounded from below, as in previous cases, so we performed intensive
simulations for assessing its properties; the acceptance rate of Algorithm
T seems to converges to one for limiting values, say $a_{1}\rightarrow-\infty$
but with $b_{1}-a_{1}$ and $\rho$ kept fixed; and to be bounded
from below by $1/2.$

\subsection{Numerical illustration}

\begin{figure}
\begin{centering}
\includegraphics[scale=0.55]{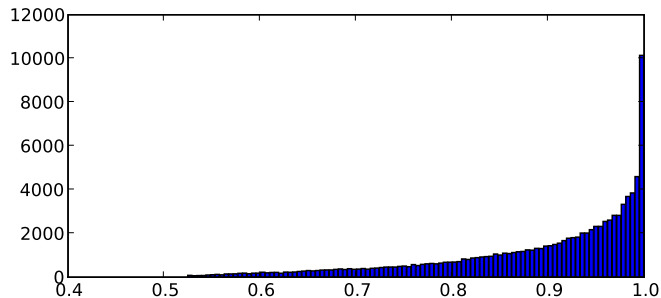}
\par\end{centering}

\caption{\label{fig:histarab}Histogram of acceptance rates for $10^{5}$ simulated
parameters $(\rho,a_{1},b_{1},a_{2},b_{2})$, with $\rho\sim U[-1,1]$,
$a_{1}$, $a_{2}\sim N(0,2^{2})$, and $b_{i}=a_{i}+2e_{i}$ with
$e_{i}\sim\mathrm{Exp}(1)$, $i=1,2.$ }

\end{figure}

We simulated $10^{5}$ parameters $(a_{1},b_{1},a_{2},b_{2},\rho)$,
where $\rho\sim U[-1,1]$, $a_{1}$, $a_{2}\sim N(0,2^{2})$, and
$b_{i}=a_{i}+2e_{i}$ with $e_{i}\sim\mathrm{Exp}(1)$, $i=1,2$.
For each parameter, we evaluated the acceptance rate of our algorithm
by computing the average acceptance probability over the proposed
draws generated by $1000$ runs. Figure \ref{fig:histarab} reports
the histogram of the acceptance rates. About $90\%$ of these values
are above $0.71$, about $99\%$ are above $0.55$, and all values
are above $0.47$, as expected. 

\section{Generalisation to 3 or more dimensions}

We discuss briefly the problem of simulating $d$-dimensional truncated
Gaussian distributions, for $d\geq3$ and for semi-finite truncation
intervals; i.e., $X\sim N_{d}(0_{d},\Sigma)$, subject to $X_{i}\geq a_{i}$,
$i=1,\ldots,d$. As we have done for $d=2$, we assume without loss
of generality that $a_{1}\geq\ldots\geq a_{d}$. It does not seem
possible to generalise to dimension $d\geq3$ algorithms based on
mixture proposals, i.e., $M^{+}$ and $M^{-}$, as the corresponding
mixture weights would involve intractable integrals. But algorithms
$S^{+}$ and $S^{-}$ can be generalised to larger dimensions, as
explained below, which makes it possible to sample $X$ under certain
conditions on $\Sigma$ and $a=(a_{1},\ldots,a_{d}$).

\subsection{Extension of Algorithm $S^{+}$}

An obvious way of generalising Algorithm $S^{+}$ to 3 dimensions
is to do the following. Let $Q=(q_{ij})=\Sigma^{-1}$ and $\Sigma_{12}$
denote the sub-matrix obtained by removing the last row and the last
column from $\Sigma$, then: 
\begin{enumerate}
\item Sample $(X_{1},X_{2})\sim N_{2}(0_{2},\Sigma_{12})$ conditional on
$X_{1}\geq a_{1}$ and $X_{2}\geq a_{2}$ (using one of the algorithms
discussed in Section \ref{sec:bitgsemi}) 
\item Sample from the unconstrained conditional distribution of $X_{3}$:
\[
X_{3}|\left\{ X_{1}=x_{1},X_{2}=x_{2}\right\} \sim N\left(-\frac{q_{13}x_{1}+q_{23}x_{2}}{q_{33}},\frac{1}{q_{33}}\right)\]

\item If $x_{3}\geq a_{3}$, accept the simulated vector $(x_{1},x_{2},x_{3})$;
otherwise reject and go to Step 1. 
\end{enumerate}
One easily shows that the acceptance rate corresponding to Step 3
is larger than $1/2$ under the following set of conditions: $ $$q_{13}\leq0$,
$q_{23}\leq0$ and $q_{13}a_{1}+q_{23}a_{2}+q_{33}a_{3}\leq0$. 

One may iterate the above principle so as to extend Algorithm $S^{+}$
to any dimension $d$; i.e., for $d=4$, add Step 4 where $X_{4}$
is simulated from the appropriate conditional distribution and accept
if $X_{4}\geq a_{4}$; provided appropriate conditions similar to
those above, are iteratively verified, one obtains an overall acceptance
rate that is at least $2^{-(d-1)}$, since each rejection step (including
Step 1 above for the first two variates $X_{1}$ and $X_{2}$) induces
an acceptance rate that is at least $1/2$. 

We note that these iterative conditions imply in particular that any
pair of components of $X$ is positively correlated, except for $(X_{1},X_{2})$
which may have any type of correlation. There are several practical
settings where this assumption is met, such as in Gaussian Markov
random fields models \citep[e.g.][]{rue:book} where one would impose
positive correlation between neighbour nodes. Thus, in such or other
particular settings, and since the acceptance rate is expected to
be higher than its lower bound for most parameters, as observed in
two dimensions, the above algorithm may remain practical for dimensions
as large as 4 or 5. For instance, in a Markov chain Monte Carlo context
involving truncated Gaussian vectors or large dimension, one may try
to form a larger and larger block, by including one variable at a
time, checking the recursive assumptions above, and stop when either
they are no longer met or the acceptance rate is too small.

\subsection{Extension of Algorithm $S^{-}$}

Again, assuming $d=3$, one notes that the marginal distribution of
$(X_{1},X_{2})$ is:
\[
p(x_{1},x_{2})\propto\exp\left\{
  -\frac{1}{2}\sum_{i,j=1}^{2}q_{ij}x_{i}x_{j}\right\}
\Phi\left(-\frac{\sum_{i=1}^{2}q_{i3}x_{i}+q_{33}a_{3}}{q_{33}^{1/2}}\right)I\left(x_{1}\geq
  a_{1};x_{2}\geq a_{2}\right)\] 
which suggests the following
bivariate truncated Gaussian density as a proposal density:
 \[
p(x_{1},x_{2})\propto\exp\left\{
  -\frac{1}{2}\sum_{i,j=1}^{2}q_{ij}x_{i}x_{j}-\frac{\left(\sum_{i=1}^{2}q_{i3}x_{i}+q_{33}a_{3}\right)^{2}}{2q_{33}}\right\}
I\left(x_{1}\geq a_{1};x_{2}\geq a_{2}\right)\] 
based on inequality
(\ref{eq:inm}). For given $x_{1}$ and $x_{2}$, the acceptance
probability is therefore:
\[
\psi\left(\frac{\sum_{i=1}^{2}q_{i3}x_{i}+q_{33}a_{3}}{q_{33}^{1/2}}\right)
/c\left(\frac{\sum_{i=1}^{2}q_{i3}x_{i}+q_{33}a_{3}}{q_{33}^{1/2}}\right)
\]
where we recall that $\psi(x)=\Phi(-x)/\varphi(x)$. Using the same
type of calculations as in Section \ref{sub:Case-Mp}, one may show
that the expectation of the acceptance probability above is larger
than or equal to $1/2$ provided $q_{13}\geq0$, $q_{23}\geq0$, $ $and
$q_{13}a_{1}+q_{23}a_{2}+q_{33}a_{3}\geq0$. Again, this means that the
overall acceptance rate is larger than or equal to $1/4$.

As in the previous section, one may iterate the construction above,
so as to obtain a simulation algorithm for any dimension $d$, the
acceptance rate of which is bounded from below by $2^{-(d-1)}$. This
requires checking recursively conditions similar to those above. The
same remarks in the previous subsection relative to the applicability
of this algorithm may be repeated here.

\section{Conclusion}

We focused in this paper on the simulation of independent truncated
Gaussian variables, but similar ideas can be used in other settings,
such as importance sampling or MCMC. For instance, in case $T$, see
Section 
\ref{sub:Case-T}, one may use the derived Gaussian approximation
as an importance distribution, rather than a basis of an ARS algorithm.
The same remark applies to most of our algorithms. 
% In particular,
% an interesting direction for further research is the calculation of
% multivariate Gaussian probabilities, see e.g. \citet{genz1992ncm}
% and \citet{craig2008}. 

As briefly mentioned in the previous section, if one needs to simulate
a vector from a high-dimensional truncated Gaussian distribution using
MCMC, one may ask how to choose blocks of two or more variables, which
will be updated using the algorithms proposed in this paper, in a
way that ensures good MCMC convergence properties. A good strategy
would be to form a first block of two variables with strong (conditional)
correlation, then to see if additional variables may be included in
that block, using the conditions given in the previous section, and
repeat this process until all variables are included in a block of
at least two variables. But more research is required to find the
best trade-off in terms of convenience and efficiency. 

Open source programs implementing the proposed algorithms are available
at the author's personal page on the website of his
institution,  \url{www.crest.fr}.

\section*{Acknowledgements}

The author thanks Paul Fearnhead, Pierre L'Ecuyer, Christian Robert,
C\^{o}me Roero, H\aa{}vard Rue, Florian Pelgrin, and the referees for helpful
comments.

\bibliographystyle{apalike}
\bibliography{complete}

\newpage

\section*{Appendix A: Outline of the univariate algorithm for a
  semi-finite truncation interval}

Note $ $\texttt{Devroye($a$)} refers to Devroye's algorithm, $
$
\texttt{Direct($a$)} refers to the rejection algorithm based on the
non truncated Gaussian distribution, see Section
\ref{sec:revi-curr-algor} for details. In both cases the input $a$ is
the truncation point. Pre-computed constants consist of five
floating-point tables: $(x_{i}),$ $(y_{i})$, $(\underline{y}_{i})$,
$(d_{i})$ and $(\delta_{i})$; one integer table: $(j_{k})$, plus two
design parameters $a_{\min}$, and $a_{\max}$. 

\bigskip

\begin{algorithm}[H]
%\caption{Univariate algorithm for sampling $TN_{[a,+\infty)}(0,1)$}
\begin{algorithmic}
\REQUIRE  $a$ \COMMENT{truncation point}
\ENSURE  $x$  \COMMENT{simulated value}
\IF {$a<a_{\min}$}
   \RETURN Direct($a$)
\ELSIF {$a>a_{\max}$}
   \RETURN Devroye($a$)
\ENDIF
\STATE $i_{a}\gets j_{\left\lfloor a/h\right\rfloor }$
\LOOP
    \STATE Sample integer $i$ uniformly between~$i_{a}$~and~$N$
    \IF[rightmost region] {$i=N$}
        \RETURN Devroye($x_N$) 
    \ELSIF[two leftmost regions] {$i\leq i_a+1$}
        \STATE Sample $u\sim U[0,1]$
        \STATE $x=x_{i}+d_{i}*u$
        \IF {$x\geq a$}
            \STATE Sample $v\sim U[0,1]$
            \STATE $y\gets y_i*v$
            \IF {$y\leq \underline{y}_i$}
                \RETURN $x$
            \ELSIF {$y\leq \varphi(x)$}
                \RETURN $x$
            \ENDIF
        \ENDIF
    \ELSE[all the other regions]   
        \STATE Sample $u\sim U[0,1]$
        \STATE {$y\gets u * y_i$}
        \IF[occurs with high probability]  {$y\leq \underline{y}_i$}
            \RETURN $x_i+u * \delta_i$
        \ELSE
            \STATE Sample $v\sim U[0,1]$
            \STATE $x\gets x_i+d_i*v$
            \IF {$y\leq\varphi(x)$}
                \RETURN $x$
            \ENDIF 
        \ENDIF
    \ENDIF 
\ENDLOOP
\end{algorithmic}
\end{algorithm}
%
% \begin{algorithm}[H]
% \begin{lyxcode}

% ~~~\textbf{return}~Direct($a$)

% \textbf{if}~$a$>$a_{\max}$~~~~~

% ~~~\textbf{return}~Devroye($a$)~

% \textbf{set}~$i_{a}=j_{\left\lfloor a/h\right\rfloor }$~

% \textbf{Label}~0~

% \textbf{sample}~integer~$i$~uniformly~between~$i_{a}$~and~$N$

% \textbf{sample}~$u\sim U[0,1]$

% if~$i=i_{a}$~~~~~~\emph{~~~~~~~~~~~~\%leftmost~region}

% ~~~~set~$x=x_{i}+d_{i}*u$

% ~~~~if~$x<a$

% ~~~~~~~go~to~Label~0

% ~~~~sample~$v\sim N(0,1)$

% ~~~~set~$y=y_{i}*v$

% ~~~~if~$y<\underline{y}_{i}$

% ~~~~~~~~return~x

% ~~~~if~$y\leq\varphi(x)$

% ~~~~~~~~return~x

% ~~~~go~to~Label~0

% \textbf{if}~$i=N+1$~~~\emph{~~~~~~~}

% ~~\textbf{return}~Devroye($x_{N}$)~~~~\emph{~\%right~tail}

% \textbf{set}~$y=u*y_{i}$

% \textbf{if}~$y<\underline{y}_{i}$~~~~\emph{~}

% ~~~~\textbf{return}~$x_{i}+u*d_{i}$~~~~~~~\emph{\%occurs~with~high~probability}

% \textbf{else}

% ~~~~\textbf{sample}~$v\sim U[0,1]$

% ~~~~\textbf{set}~$x=x_{i}+d_{i}*v$

% ~~~~\textbf{if}~$y<\varphi(x)$

% ~~~~~~~~\textbf{return}~$x$

% ~~~~\textbf{else}~

% ~~~~~~~~\textbf{go~to}~Label~0
% \end{lyxcode}

% \end{algorithm}

\section*{Appendix B: Lower bounds for Acceptance rates }

\section*{B1. algorithm $S^{-}$}

Let $A(a_{1},a_{2},\rho)$ the acceptance rate (\ref{eq:arsm}), which
we rewrite as:\[
A(a_{1},a_{2},\rho)=E_{TN_{[\alpha,\infty)}(\beta,\rho^{2})}\left[\psi\left(Z\right)\right]/c\left(-\alpha\right)\]
where $Z=-(\rho X_{1}-a_{2})/\nu\sim TN_{[\alpha,\infty)}(\beta,\rho^{2})$,
$\alpha=(a_{2}-\rho a_{1})/\nu$, and $\beta=a_{2}\nu$. Note that
$\alpha\geq0$, $\beta\leq\alpha$, and $\psi$ is a decreasing function.
Thus, the quantity above is a decreasing function of $\beta$. (To
see this, one can rewrite the distribution of $Z$ as \[
Z=\beta+\rho\Phi^{-1}\left(U+(1-U)\Phi(\frac{\alpha-\beta}{\rho})\right)\]
where $U$ is an uniform variate, and check that, conditional on $U=u$,
$Z$ is a decreasing function of $\beta$.). Thus, the above quantity
is larger than or equal to the same quantity, but with $\beta=\alpha$:
\[
A(a_{1},a_{2},\rho)\geq E_{TN_{[\alpha,\infty)}(\alpha,\rho^{2})}\left[\psi\left(Z\right)\right]/c\left(-\alpha\right)=E_{TN_{[0,\infty)}(0,1)}\left[\psi\left(\rho Z'+\alpha\right)\right]/c\left(-\alpha\right)\]
which is a decreasing function of $\rho,$ hence \begin{equation}
A(a_{1},a_{2},\rho)\geq E_{TN_{[0,\infty)}(0,1)}\left[\psi\left(Z'+\alpha\right)\right]/c\left(-\alpha\right),\label{eq:lowbounda}\end{equation}
and since $c(-\alpha)=(\sqrt{\pi/2})\wedge(1/\alpha)$, one can show
that the bound is minimised for $\alpha=\sqrt{\pi/2}$, which leads
to:\[
A(a_{1},a_{2},\rho)\geq\sqrt{\frac{2}{\pi}}E_{TN_{[0,\infty)}(0,1)}\left[\psi\left(Z'+\sqrt{\frac{2}{\pi}}\right)\right]\approx0.416.\]

This lower bound is not sharp, because not all combinations of $(\alpha,\beta,\rho)$
are valid, even in the constraints $\alpha\geq0$, $\beta\leq\alpha$
are taken into account; for instance, $\beta=\alpha$ implies that
$a_{1}=\rho a_{2}\leq\rho^{2}a_{1}$, which is impossible if $\rho\neq1.$
Our simulations suggests that the optimal lower bound is $1/2$, see
Section \ref{sub:Results}.

\section*{B2. algorithm $M^{+}$}

One easily shows that $\chi(x)/\chi(x')\geq1/2$ for all $x,x'\in[0,x_{d}]$,
with $x_{d}\approx3.117$. Thus, if $(a_{2}-\rho a_{1})/\nu\leq x_{d}$,
the acceptance rate is larger than or equal to $1/2$ by construction.
Now assume that $(a_{2}-\rho a_{1})/\nu>x_{d}$; note that $\chi(x)$
is an increasing function for $x>x_{d}$. Since $a_{1}\geq a_{2}$
and $\rho\leq1$, one has \[
\theta=\rho(a_{2}+\lambda\nu)\leq a_{2}+\nu(\lambda\rho-x_{d})<a_{1}.\]

The acceptance rate equals \begin{equation}
  E_{TN_{[a_{1},a_{2}/\rho]}(\theta,\nu^{2})}\left[\frac{\chi\left(\frac{a_{2}-\rho
          X_{1}}{\nu}\right)}{d\left(\frac{a_{2}-\rho
          a_{1}}{\nu}\right)}\right]=E_{TN_{[0,z_{\max}]}(\eta,\rho^{2})}\left[\frac{\chi\left(Z\right)}{\chi\left(z_{\max}\right)}\right]\label{eq:armp}\end{equation}
where $z_{\max}=(a_{2}-\rho a_{1})/\nu$, and
$\eta=(a_{2}-\rho\theta)/\nu>z_{\max}$; note
$d(z_{\max})=\chi(z_{\max})$ provided $z_{\max}>0.751$, but we assumed
that $z_{max}>x_{d}\approx3.117$. The $TN_{[0,(a_{2}-\rho
  a_{1})/\nu]}(\eta,\rho^{2})$ distribution should concentrate its
mass at the right edge of interval $[0,z_{\max}]$, and
$\chi(Z)/\chi(z_{\max})$ should take values close to
one. Specifically, one has that $z\Phi(-z)/\varphi(z)\in(0.84,1)$ for
$z>2$, thus \[
\frac{\chi(z)}{\chi(z_{\max})}>0.84e^{\lambda(z-z_{\max})}\geq0.5\]
for $z\in[z_{\max}-0.76,z_{\max}]$. Therefore (\ref{eq:armp}) is
larger than $0.5$ times the probability that $Z\geq z_{\max}-0.76$,
for $Z\sim TN_{[0,z_{\max}]}(\eta,\rho^{2})$, which is larger than or
equal to $0.44$, the probability of the same event with respect to
$Z\sim TN_{[0,z_{\max}]}(z_{\max},1)$, for $z_{\max}$. This gives $ $a
lower bound for (\ref{eq:armp}) of $0.22$. We obtained sharper bounds
with more tedious calculations (omitted here), but more importantly,
our simulation studies indicates that the optimal lower bound is
likely to be larger than or equal to $1/2$.
\end{document}